\documentclass[aps,superscriptaddress,twocolumn,nofootinbib]{revtex4}

\usepackage{graphicx}
\usepackage{bm}
\usepackage{amsmath,epsfig}

\begin{document}



\title{The Tenth Article of Ettore Majorana}

\author{Rosario Nunzio Mantegna}
\affiliation{Dipartimento di Fisica e Tecnologie Relative, 
Universit\`a di Palermo, Viale delle Scienze, I-90128 Palermo, Italy}


\begin{abstract}
This year is the centenary of the birth of Ettore Majorana, one of the major Italian physicists of all times. In this note we briefly sketch a few biographical details about Ettore Majorana and introduce and discuss the main points of Majorana's 10th article. In his article Majorana explicitly considers quantum mechanics as an irreducible statistical theory because the theory is not able to describe the time evolution of a single particle or atom in a precise environment at a deterministic level. This lack of determinism at the level of an elementary physical system motivated him to suggest a formal analogy between statistical laws observed in physics and in the social sciences. We hope the occasion of the centenary of the birth of Ettore Majorana will be useful to remember and to reconsider not only his exceptional achievements in theoretical physics but also his fresh and original views on the role of statistical laws in physics and in other disciplines such as the social sciences.
 \end{abstract}

%
\maketitle

This year is the centenary of the birth of Ettore Majorana, one of the major Italian physicists of all times. The centenary has triggered a series of initiatives. Among them there is the International Conference on "Ettore Majorana's legacy and the Physics of the XXI century" organized by the Dipartimento di Fisica e Astronomia of University of Catania, Italy that will be held in Catania on October 5-6. Another initiative of the Società Italiana di Fisica concerns the publication of a volume including all Majorana's articles both in the original language (in most cases Italian) and in English. The volume will also contain a brief descriptive note on each of the articles. Lastly, the International School on Complexity of the "Ettore Majorana Foundation and Centre for Scientific Culture" of Erice, Italy co-directed by A. Zichichi, G. Benedek, M. Gell-Mann, L. Pietronero and C. Tsallis, has scheduled a Course on September 17-23 entitled "Statistical Laws in Physics and Economics" directly inspired by one of Majorana's articles, specifically his 10th article. This article is probably the least known among his articles. 

In this note we briefly sketch a few biographical details about Ettore Majorana and introduce and discuss the main points of Majorana's 10th article. The biographical notes are mainly based on information that can be found in \cite{Amaldi,Recami}. The aim of this note is to invite the readership of Europhysics News to consider the main message of Majorana's last contribution with its focus on the value of statistical laws in physics and social sciences \cite{Majorana}.

Ettore Majorana was born in Catania, Italy on August 5th 1906 in a well known and influential family of that city. His father was Fabio Massimo Majorana, an engineer that was appointed to be General Inspector of the Italian Ministry of Communication in Rome in 1928. Fabio Massimo Majorana was the younger brother of Quirino Majorana a renowned professor of experimental physics at Bologna University and a member of the prestigious Accademia dei Lincei. Fabio Massimo Majorana married Dorina Corso and they had two daughters and three sons, specifically, Rosina, Salvatore, Luciano, Ettore and Maria.

The family moved to Rome in 1921. Ettore Majorana studied at the secondary schools "Istituto Massimo" and "Liceo Statale Torquato Tasso" where he obtained his diploma. He then enrolled at the School of Engineering of Rome University (at that time only "La Sapienza" University existed). At the School of Engineering he started to interact with Emilio Segré. In 1926 Emilio Segré moved from the School of Engineering to the Institute of Physics where Enrico Fermi (at that time 26 years old) was nominated to the position of full professor of Theoretical Physics. The same path was followed by other successful Italian physicists such as, for example, Edoardo Amaldi. In the new environment of physics, Emilio Segré was often talking about the exceptional qualities of Ettore Majorana and meanwhile he attempted to convince Ettore to join the group of physicists led by Enrico Fermi. The passage of Ettore Majorana from the School of Engineering to the Institute of Physics did happen at the beginning of 1928. At that time the involvement of Ettore Majorana as a student and then as a professional physicist started. 

Ettore Majorana was one of the greatest theoretical physicists of the heroic period of the development of quantum mechanics and nuclear physics in the first half of the last century. Ettore Majorana graduated in physics in 1929 under the supervision of E. Fermi. During the years from 1929 to 1933 he devoted all his energy to theoretical physics and produced most of his work in this field. In 1933 he visited Lipsia, where he positively interact with W. Heisenberg, and briefly Copenhagen where he met Niels Bohr. After his return to Rome during the fall of 1933 his involvement in theoretical physics research declined until 1937 when he again showed an active interest in theoretical physics by taking part in a national examination to obtain a position of professor of theoretical physics at Palermo University. The Committee composed of E. Fermi, O. Lazzarino, E. Persico, G. Polvani and A. Carelli, unanimously considered his work to be outstanding and above comparison with the work of the other candidates and therefore proposed that the Minister of Education should nominate E. Majorana as a full professor for "chiara fama", namely outside the examination procedure. He therefore became professor of theoretical physics at Naples University where he taught a quantum mechanics course during the academic year 1937-38. 

Ettore Majorana mysteriously disappeared in March 1938 during a trip on the ship connecting Palermo to Naples. His body was never found although the search for Majorana both by government officials and members of the Majorana family continued for a long time. Enrico Fermi had a very high estimation of him. In the letter he wrote in 1938 to the Italian prime Minister of that time, Benito Mussolini, asking the government to intensify the support of Ettore's research, Fermi stated "I have no hesitation to state to You, and I am not saying this as an hyperbolic statement, that of all Italian and foreign scholars that I have had the opportunity to meet, Majorana is among all of them the one that has most struck me for his deep sharpness" \cite{Recami}.

Majorana was not a prolific author. He just published 9 articles before his disappearance and a 10th article, whose manuscript was found by Majorana's brother among his files, was published in 1942 after his disappearance in the international Italian journal Scientia, through the interest of his friend Giovanni Gentile Jr.. Nine of these articles were written in Italian and one in German. Italian is not a widespread language and this limitation has prevented Majorana's work becoming known and correctly evaluated by a vast community of scientists. I have translated the 10th article of Majorana "The value of statistical laws in physics and social sciences" to provide to a broad audience of physicists the possibility of direct access to the article. The translation was recently published in Quantitative Finance 5, 133-140 (2005). 

The article is a rather special article in several respects. In the original presentation for Scientia, Giovanni Gentile Jr. wrote that the article was originally written for a sociology journal. This article was therefore intended to present the point of view of a physicist about the value of statistical laws in physics and social sciences to scholars of a broad spectrum of different disciplines such as sociology and economics. In his article, Majorana considers quantum mechanics as a fundamental and successful theory able to describe the basic processes involving single particles and atoms. He explicitly considers the theory as a statistical theory because the theory is not able to describe the time evolution of a single particle or atom in a precise environment at a deterministic level. As an example of the lack of determinism in the time evolution of a single system he discusses the case of the decay of a radioactive atom. This lack of determinism at the level of an elementary physical system motivated him to suggest a formal analogy between statistical laws observed in physics and the social sciences. In his article, he concludes that there is an "essential analogy between physics and the social sciences, between which an identity of value and method has turned out". These words seem to pioneer the view that an investigation of complex systems (indeed this term is literally present in the article) of economic or social origin might be conducted on the same epistemological basis as the modeling of physical systems.

His conclusion was considered as rather peculiar and was accepted as a general belief by only a minority of physicists for several years. For example in the book "La vita e l'opera di Ettore Majorana" \cite{Amaldi}, Edoardo Amaldi just wrote a single sentence on this article in a biographical and scientific note of 49 pages. Even today, Majorana's point of view might indeed still be rather unpopular among mainstream physicists, in spite of more than 70 years of quantum mechanics and after some major breakthrough in the fields of critical phenomena and chaos theory. 

There is a pioneering nature of this article both from the perspective of physics and economics. From the physics point of view, Majorana took a clear position about the key aspect that quantum mechanics forces scientists to use a statistical description down to events involving single entities. From the point of view of economics and social sciences, there is an emphasis on the observation that statistical laws have to be used in economic and social modeling. It should be noted this position was not that of the majority of scholars working in the thirties of the last century in both the disciplines considered. In fact, during the thirties of the last century the interaction between social sciences and natural scientists was developed under the paradigm of celestial mechanics (the only exception to this approach was the one pursued by Louis Bachelier that at that time had no impact on the academy \cite{Bachelier}). This interaction goes back to the development of general equilibrium theory pursued by Walras, Pareto, Schlesinger and Wald. The emphasis of Majorana on the intrinsic statistical nature of most of the underlying processes describing natural phenomena suggests that statistical laws should be incorporated into a modeling approach to social phenomena. This approach has eventually found its best achievement in finance with the Black and Scholes modeling of option pricing \cite{BlackScholes}.

The topic considered by Majorana in his article is timely today for a series of reasons. First, it should be noted that a cross-disciplinary consensus about the epistemological value of statistical laws in different disciplines is not easily found today. The major paradigm of the validity of a scientific theory is still based on the falsification procedure of a law. It is undisputable that this approach has been devised having in mind the most characteristic laws in physical sciences, i.e. deterministic laws or laws having a deterministic part (as is the case for quantum mechanics when the time evolution of the wave function is considered or for random walk theory when one considers the statistical description of an ensemble of walks). During the past years it has been progressively realized that such an approach might not be the most appropriate to other disciplines such as, for example, biology. For this important and successful discipline, the nature of the laws (or sometimes theories) is often intrinsically related to the prevalence of indeterminacy owing to the high frequency of stochastic processes unavoidably involved and moreover a double causality (one related to the external conditions and forces and one governed by the amount of information inherited at the biological level under consideration) is present in most cases \cite{Mayr}. Similar observations are most probably also valid for social sciences. 

Physics might certainly benefit from a deeper understanding of the role, necessity and peculiarity of statistical laws in physics. Some of the statistical laws are eventually reinterpreted in terms of more fundamental and deterministic laws. However there are cases when a reduction seems to be impossible. One of these cases is indeed quantum mechanics and other more recent examples concern the topics of chaos and critical phenomena theory. By somewhat inverting the perspective of the relations between physics, biology and the social sciences, it might be worth discussing the possibility that physicists should also start to consider as proper to their discipline the investigation of systems where the basic elements composing the system are in possession of a certain level of internal information and are characterized by a certain ability to react to external stimuli by properly processing the available external information with inherited or adaptive rules. Investigations of this kind have been performed with tools of statistical physics properly adapted or extended. For example, one of these models is the minority game \cite{Challet,Coolen} recently investigated within the new research field of econophysics \cite{Farmer}. 

In summary, the 10th article of Majorana raised the necessity of focusing the attention of several disciplines on the value and nature of statistical laws. From physics, to biology and to social sciences, all the scientific disciplines present statistical laws and scholars of these disciplines need to reflect about their role within each discipline. Majorana noticed that quantum mechanics made clear that a scientific description without statistical laws is impossible. Today there is still a need to assess the status of statistical laws and to consider the validation procedures that are most appropriate to these sorts of laws. Validation procedures probably need to be different from those originally devised having in mind deterministic laws. 

I hope the occasion of the centenary of the birth of Ettore Majorana will be useful to remember and to reconsider not only his exceptional achievements in theoretical physics but also his fresh and original views on the essential aspects, importance and role of statistical laws in physics and in other disciplines such as the social sciences.


\end{document}